\documentclass[%
mathleft,%
]{an}
\usepackage{graphicx}
\usepackage{times}
\begin{document}

\Pagespan{1}{}
\Yearpublication{2011}%
\Yearsubmission{2011}%
\Month{1}%
\Volume{999}%
\Issue{92}%

\title{Asiago Supernova classification program: blowing out the first two hundred candles}

\author{L. Tomasella\inst{1}\fnmsep\thanks{Corresponding author:
  \email{lina.tomasella@oapd.inaf.it} }
\and S. Benetti \inst{1}
\and E. Cappellaro  \inst{1}
\and A. Pastorello \inst{1}
\and M. Turatto  \inst{1}
\and R. Barbon \inst{1}
\and N. Elias-Rosa  \inst{1}
\and A. Harutyunyan \inst{2}
\and  P. Ochner  \inst{1}
\and  L. Tartaglia  \inst{1}
\and S.Valenti  \inst{3,4}
}
\titlerunning{Asiago SN classification Program}
\authorrunning{L. Tomasella}
\institute{
$^{1}$INAF, Osservatorio Astronomico di Padova, 35122 Padova, Italy\\
$^{2}$Fundaci\'on Galileo Galilei, INAF Telescopio Nazionale Galileo, Rambla Jos\'e Ana Fern\'andez P\'erez 7, 38712 Bre\~na Baja, TF, Spain\\
$^{3}$Department of Physics, University of California, Santa Barbara, Broida Hall, Mail Code 9530, Santa Barbara, CA 93106-9530, USA\\
$^{4}$Las Cumbres Observatory Global Telescope Network, 6740 Cortona Dr., Suite 102, Goleta, CA 93117, USA}


\abstract{We present the compilation of the first 221 supernovae classified during the Asiago Classification Program (ACP). The details of transients classification and the preliminarily reduced spectra, in fits format, are immediately posted on the Padova-Asiago SN group web site. The achieved performances for the first  2 years of the ACP are analysed, showing that half of all our classifications were made within 5 days from transient detection. 
The distribution of the supernova types of this sample resembles the distribution of the general list of all the supernovae listed in the Asiago SN catalog (ASNC$^8$, Barbon et al. 1999). Finally, we use our sub-sample of 78 core-collapse supernovae, for which we retrieve the host-galaxy morphology and $r$-band absolute magnitudes, to study the observed subtype distribution in dwarf compared to giant galaxies. This ongoing program will give its contribution to the classification of the large number of transients that will be soon delivered by the Gaia mission. 
}

\maketitle

\section{Introduction}

The surveillance of the transient sky has greatly improved in the last ten years with the contribution of a growing number of surveys. Panoramic surveys of the nearby Universe like the Catalina Real-Time transient Survey (CRTS)\footnote{http://crts.caltech.edu}, the Palomar Transient Factory (PTF)\footnote{http://ptf.caltech.edu/iptf/}, the La Silla Quest\footnote{http://hep.yale.edu/lasillaquest}, the Mobile Astronomical System of theTElescope-Robots (MASTER)\footnote{http://observ.pereplet.ru}, the Optical Gravitational Lensing Experiment \linebreak (OGLE)\footnote{http://ogle.astrouw.edu.pl}, etc, have boosted the number of supernova (SN) discoveries from $\sim200$ in 2000 to $\sim1000$ in 2013. Many of these surveys post their discoveries in real time, allowing for a change of approach in SN science: for a specific SN science case the relevant events can be selected and studied.

However, to really exploit the transient search efforts, we need to be able to identify, as early as possible, the  event class with a prompt spectroscopic classification. Knowing the SN type and phase, one can activate a proper follow up campaign. The observing chain made of wide field searches, prompt classification and selective follow-ups has proved to be extremely productive as can be gathered from an \linebreak overview of the recent literature. As examples of the latest advances in SN researches, we recall the discovery of the class of super-luminous supernovae (SNe) whose explosion mechanism is still debated (Pastorello et al. 2010, Quimby et al. 2011, Gal-Yam 2012) and the real-time observations of the convulsions of massive stars on their path to explosion (Smith et al. 2013, Pastorello et al. 2013, Fraser et al. 2013, Margutti et al. 2014).

The classification and early follow-up of bright nearby SNe can be efficiently done with small/medium size telescopes. In the past few years we conducted an European Southern Observatory (ESO) Large Program on { \em Supernova Variety and Nucleosynthesis Yields} (2009-2012) with the ESO-NTT (New Technology Telescope), complemented \linebreak with the INAF-TNG (Telescopio Nazionale Galileo of Istituto Nazionale di Astrofisica)  for the Northern hemisphere, for  the study of selected SNe  (eg. Taubenberger et al. 2011, Patat et al. 2011, Fraser et al. 2011, Valenti et al. 2011a, Pastorello et al. 2012, Kankare et al. 2012, Pastorello et al. 2013,  Tomasella et al. 2013a). Building on this successful experience, the ESO Large Program was merged into a new major international collaboration, the Public ESO Spectroscopic Survey of Transient Objects (PESSTO)\footnote{http://www.pessto.org/pessto/index.py}, that is using a major fraction of the time at the ESO NTT at La Silla (Chile). PESSTO started in 2012 and will be active for 4 (+1) years (Smartt et al. 2013).

Despite the efforts of this large project and those of other groups worldwide, a large fraction of transients are not spectroscopically classified (around 50\%, based on the  list of Latest Supernovae\footnote{http://www.rochesterastronomy.org/snimages/}).   
In this context, we decided to give a contribution to the classification of the brightest targets of the Northern hemisphere by exploiting our access at the observing facilities in Asiago, in particular the 1.82m Copernico telescope at Cima Ekar, operated by INAF Astronomical Observatory of Padova (OAPd). A parallel project is the photometric and spectroscopic follow-up of the most interesting transients (classified or not by us), which is not addressed in this paper. 
In fact, in the last few years the observations in Asiago have been deeply reorganised and a remarkable amount of telescope-time is allocated for two or three Large Programs. Our ongoing {\em Classification and follow-up of extragalactic transients discovered by \linebreak panoramic surveys} is one of those Large  Programs.

The Asiago SN Classification Program (ACP) started in 2011, with the aim to classify all transients that are accessible from our latitude and are bright enough for 1.82m Copernico  telescope  (apparent magnitude $\leq 19$ mag). The project is the latest evolution of Asiago SN programs, which begun in the early sixties. Among the historical achievements, we recall: the systematic SN search with Schmidt telescopes (Rosino 1964);  the first identification of peculiar type~I SNe (Bertola 1964), named Ib or Ic twenty years later (Gaskell et al. 1986); the derivation of an average light curve for type~Ia (Barbon et al. 1973) and of the (different) photometric properties of type~II (Barbon et al. 1979) SNe;  and the compilation of the Asiago SN Catalogue (Barbon et al. 1999), intended as a large database for statistical studies on the SN phenomenon (ASNC)\footnote{http://sngroup.oapd.inaf.it/asnc.html}.

\begin{figure}
\includegraphics[width=\linewidth]{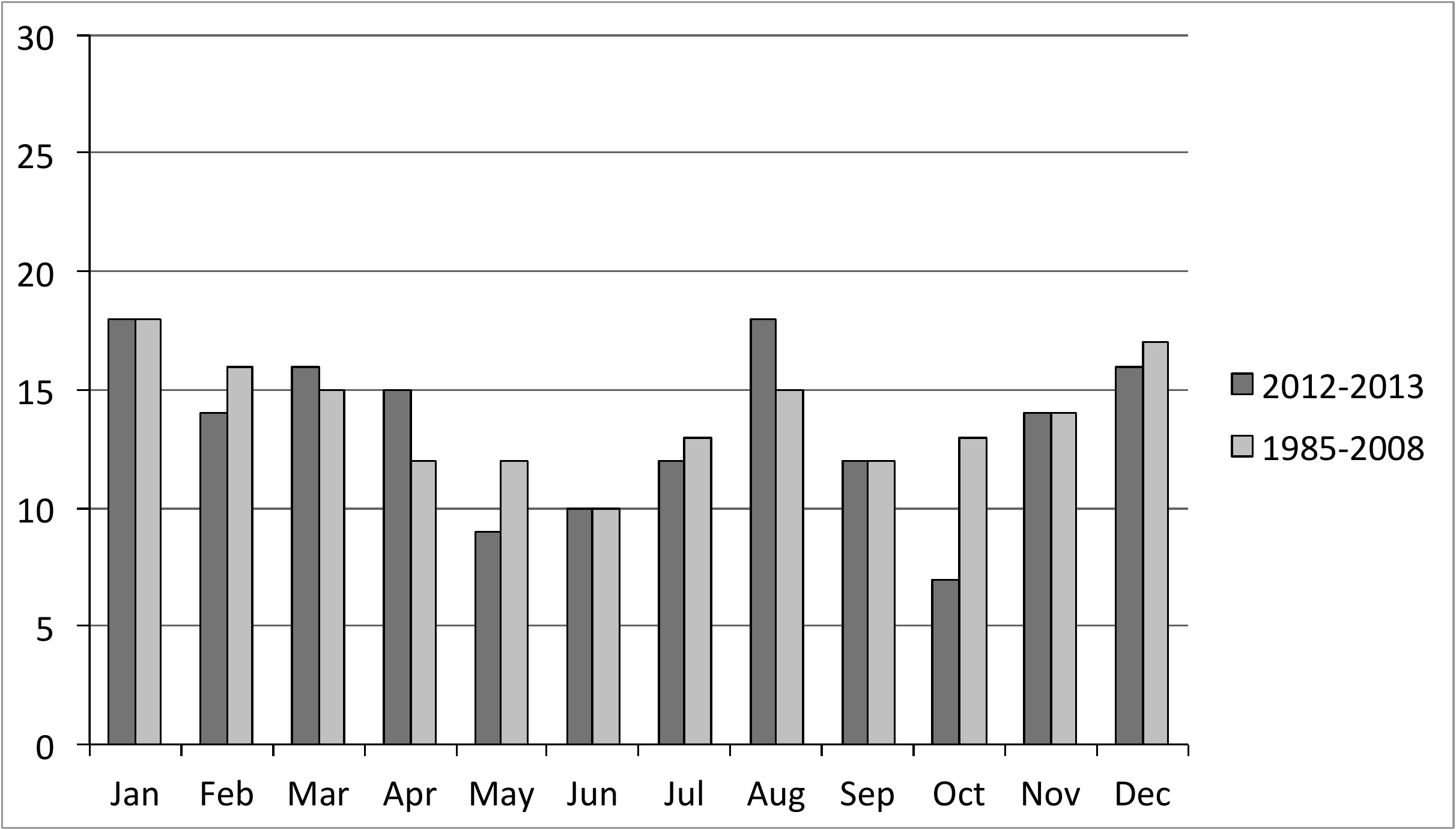}
\caption{Statistics of the open-dome nights at the Copernico 1.82m telescope (including the partially used nights) as a function of the month for the past two years (2012$-$2013) and comparison with the statistic over the past 24 years (1985-2008). June and July are mainly used for telescope and dome maintenance, tests and outreach activities. }
\label{nights}
\end{figure}

The ACP is allocated on average one week of observing nights per month at the Copernico 1.82m telescope. Typically half of the allocated time is used for classification of new targets and half to contribute to follow-up observations of selected objects. When the Copernico telescope is not available, the 1.22m Galileo telescope is used for transients  brighter than magnitude 17. 
Both these telescopes are located in the Asiago Plateau, North-East of Italy, about one hundred km from Padova, at an altitude of 1.366 m for Copernico (Mt. Ekar, $11^\circ\,34'\,08.42\,{\rm E}$, $+45^\circ\,50'\,54.52''\,{\rm N}$) and 1.045 m for Galileo ($11^\circ\,31'\,3''\,{\rm E}$, $+45^\circ\,51'\,59''\,{\rm N}$). \linebreak This site is characterised by a continental climate,  with dry winters and rainy spring and autumn. Summer time is on average favourable for observations. A statistics of the nights per month with open dome for the last years is plotted in Figure~\ref{nights}. The seeing is quite variable during the year, with an average $\sim2''$, but nights with seeing around $1''$ are frequently registered. Despite the proximity of large cities, the collaboration with Asiago Plateau Municipalities and the Regional Agency for the environmental preserve (ARPAV)  has contributed to the decreasing of the local light pollution during the past ten years (S. Ortolani, private communication).

The ACP program proceeds as follows. Suitable targets are identified among those posted by SNe searches. Soon after acquisition, the spectra are immediately reduced through a semi-automatic data reduction pipeline (Sect.~2).  The spectra are then compared to SN templates aided by automatic SN classification codes (Sect.~3.) and the classifications are disseminated  via the IAU Central Bureau Astronomical Telegrams circulars (CBET)\footnote{http://www.cbat.eps.harvard.edu/cbet/RecentCBETs.html} and/or the Astronomical Telegram  (ATEL)\footnote{http://www.astronomerstelegram.org}. Some statistics of the program performances and the properties of the classified SN sample are reported in Sect.~4.

Following what we believe is a very fruitful trend of many new projects, we made  the results of our effort  immediately public: within a few hours from observation, the details of transient classification and the (fast-)reduced spectra (fits format) are posted in the Padova-Asiago SN group web site \footnote{http://sngroup.oapd.inaf.it/}.

\section{Transient selection, observation and data reduction}

We select the  SN candidates mainly from the IAU CBAT Transient Object Confirmation Page\footnote{http://www.cbat.eps.harvard.edu/unconf/tocp.html} and from the Astronomy Section of the Rochester Academy of Sciences$^7$. In some cases, candidates are notified directly to us by SN hunters, who often are amateur astronomers, e.g.  the Italian Supernovae Search Program (ISSP)\footnote{http://italiansupernovae.org/}. 

We prioritize the transients for classification based on the following criteria:
\begin{itemize}
\item apparent magnitude at discovery $\leq18.5-19$ mag; 
\item optimal visibility from Asiago;
\item date of discovery (recent discoveries have higher priority).
\end{itemize}

Also, in order to complement PESSTO, we give priority to Northern-most transients that are unaccessible from La Silla. Table~1 shows  the statistics of classified SNe grouped by discoverer. A large fraction of our classifications (almost 44\%) concerns nearby and bright events discovered by amateurs, with 20\% of them announced by ISSP. Among the professional searches, we retrieve a major fraction  of candidates from CRTS. 

\begin{table}
\caption {Number of classified SNe (central column) grouped by discoverer (first column). The last column reports the percentage on the total number of classified SNe (221).}
\begin{center}
\begin{tabular}{lcc}
\hline
discoverer & Number & \% \\
\hline
CRTS & 83 &  37.5\\
MASTER & 18 & 8.1 \\
LOSS & 3 & 1.4 \\
TNTS & 7 & 3.2 \\
Others (professional) & 13 & 5.9 \\
Amateurs & 97 & 43.9  \\
\hline
\end{tabular}
\end{center}

CRTS, Catalina Real-time Transiet Survey; MASTER, Mobile Astronomical System of theTElescope-Robots; LOSS, Lick Observatory Supernova Search; TNTS, THU-NAOC Transient Survey; {\it Others}, all the other professional surveys (CHASE, Xing Gao, ASAS-SN). 
\end{table}

The observations are mainly carried out at the Copernico 1.82m telescope equipped with AFOSC (the Asiago Faint Object Spectrograph and Camera).  The instrument allows for imaging (field of view of 8.9 $\times$ 8.9 arcmin, sampling 0.52 arcsec/pix, with a 2 $\times$ 2 binned CCD) and for medium to low resolution spectroscopy. There is a selection of available grisms with resolutions up to 7000 (using Volume Phase Holographic gratings).
The CCD camera is an Andor DW436-BV equipped with a E2V CCD42-40 AIMO back illuminated device.
Additional details on the instrumentation can be found in the user manuals that can be downloaded from the dedicated OAPd web pages \footnote{http://www.oapd.inaf.it/index.php/en/telescopes-and-instrumentations/}.

For the classification spectra, we generally use grism \#4 (300 gr/mm), providing a wavelength range 3500-8200 \AA\/ ($\lambda_{blaze}$=5094 \AA\/; 
dispersion=208 \AA\/ mm$^{-1}$). In combination with the $1.69''$ wide slit, the spectral resolution is 1.3 nm, as measured from the FWHM of the night sky lines.

Occasionally, if the Copernico 1.82m telescope is not available and the target is bright enough (magnitude $\leq$ 17), we use the Galileo 1.22m telescope, which is the instrument of the University of Padova and is located in the historical  site of the Astrophysical Observatory of Asiago. 
The Galileo telescope is permanently equipped with Boller $\&$ Chivens long-slit spectrograph. The CCD camera is an Andor iDus DU440 back illuminated, 2048$\times$512 pixels, sampling 1~arcsec/pix\footnote{http://www.dfa.unipd.it}. The set-up for the classification program at 1.22m makes use of the 300 gr/mm grating, covering the wavelength range 3600-7900 \AA\/ ($\lambda_{blaze}$=5000 \AA\/; dispersion=166.8 \AA\/ mm$^{-1}$). In combination with a $2''$ wide slit, the spectral resolution is 0.8 nm. 

As a rule, during observation with both telescopes, the slit is aligned to the parallactic angle when the airmass exceedes 1.5.  At that airmass, the atmospheric differential refraction at the two ends of the observed wavelength range is $\sim$1.0~arcsec (cf. Filippenko 1982). Typically, exposure times range from 20 minutes to 1 hour, depending either on the magnitude of the transient and on the weather and seeing conditions.

Fast data reduction at the Copernico 1.82m telescope is performed using a dedicated pipeline. It has been developed in a {\sc python} environment by one of us (S.V.) and executes sequentially the required standard {\sc iraf} tasks. 
In short, spectral images are bias and flat-field corrected and the SN spectrum is extracted by interactively tracing the stellar profile along the spectral direction and subtracting the sky background fitted along the slit direction. Wavelength calibration is accomplished by using archive wavelength/pixel dispersion coefficients and is then adjusted to match the expected wavelengths of strong night sky lines.
Also for flux calibration we use an archive spectral sensitivity curve regularly updated with observations of spectrophotometric standard stars. The telluric absorption correction is derived from the archive spectra of spectrophotometric standards. This explains why imperfect removal can affect the profile of the SN features near the strongest atmospheric absorptions, in particular the telluric band at $7570-7750$ \AA. 

We verified that in most cases the fast reduced spectra are calibrated within a few percent on a relative flux scale. Due to unstable weather conditions, the absolute flux calibration can not be guarantied. Therefore, if a more accurate calibration is required, the 
spectra are re-reduced using  the proper calibration observations obtained in the same night as the SN spectrum. 

\section{Spectral Classification}

\begin{figure*}
\includegraphics[width=\linewidth]{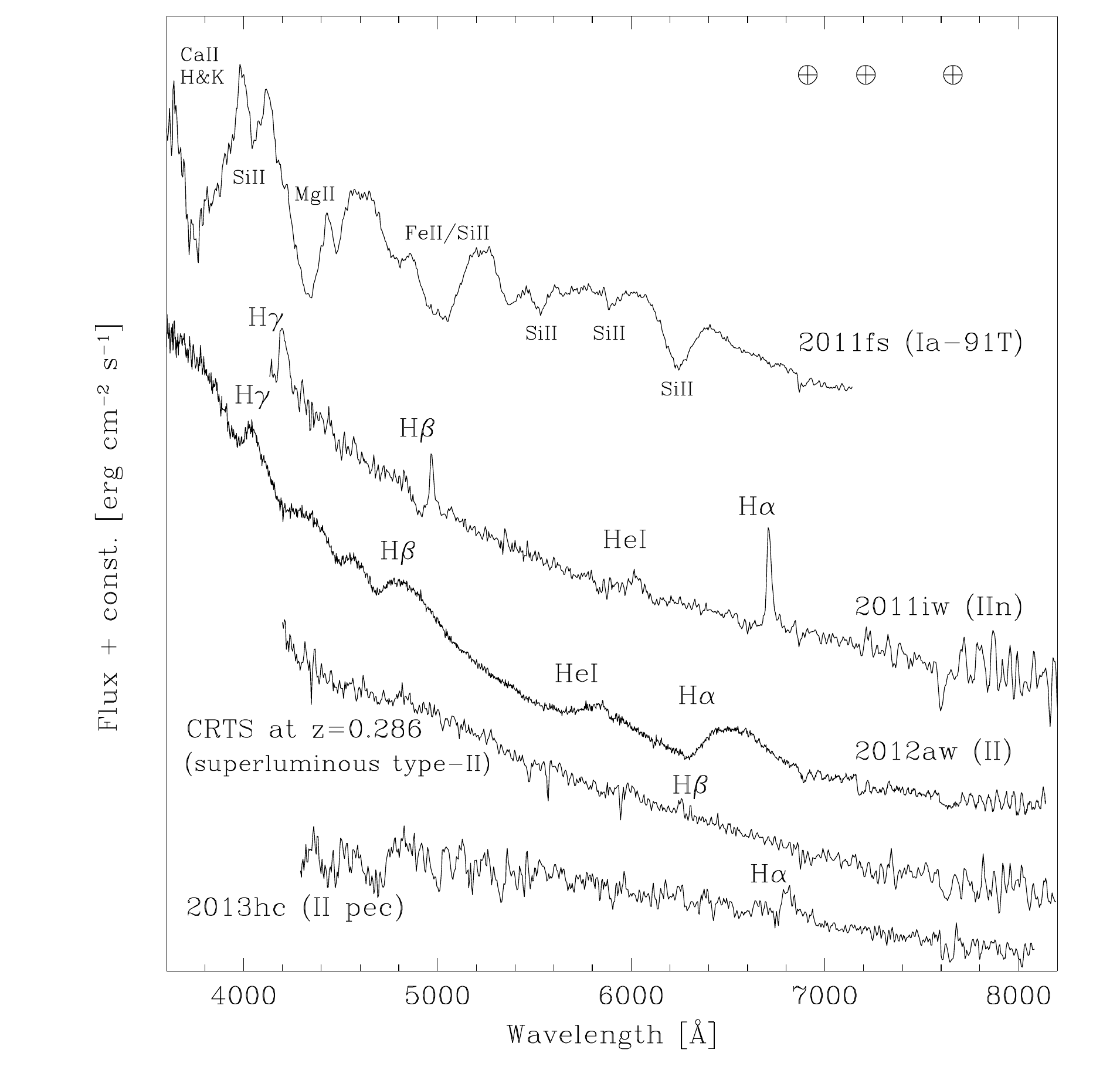}
\caption{Representative spectra of SNe classified in the framework of the ACP. From top to bottom: peculiar type-Ia SN~2011fs and spectra of core-collapse SNe 2011iw; 2012aw in M95; CSS121015:004244+132827 at z=0.286 (H$_\alpha$ is out of the wavelength range); 2013hc is the fainter classified transient (V$\sim$20 mag at classification).}
\label{spectra}
\end{figure*}

The flux calibrated spectrograms of the SN candidates are promptly examined through cross-correlation with libraries of SN template spectra using different automatic tools, \linebreak namely the  GEneric cLAssification TOol\footnote{https://gelato.tng.iac.es} ({\sc GELATO},  \linebreak Harutyunyan et al. 2008), the Supernova Identification code ({\sc SNID}; Blondin and Tonry 2007), and the {\sc SUPERFIT} code (Howell et al. 2005).  
Indeed the codes have different approaches and databases and they can have different  performances, depending on the signal-to-noise ratio of the spectra, ancillary information available (in particular redshift and reddening) and SN types.

In particular, {\sc SNID} is the most efficient algorithm if the redshift of the SN host galaxy is unknown. Moreover, it is tuned to obtain an accurate estimate of the phase for type Ia SNe. On the other hand, {\sc SUPERFIT}  is useful when the SN is deeply embedded on the host galaxy background because the possible contamination is accounted in the fit. 

{\sc GELATO} was developed by one of us (A.H., Harutyunyan et al. 2008) and it relies on an extended database of SN spectra  of all types collected by our group during over two decades, and in few cases complemented with spectra of noticeable SN types retrieved from public archives such as SUSPECT (Richardson et al. 2002)\footnote{http://nhn.nhn.ou.edu/~suspect/} and WISeREP (Yaron and Gal-Yam 2012)\footnote{http://www.weizmann.ac.il/astrophysics/wiserep/}. One of the distinctive features of the code is its web-based interface. This has two advantages: $i)$ it does not require any user installation and it is accessible from any computer through standard web browsers; $ii)$ upgrades of GELATO software or new templates additions to its database are immediately available to the users.
{\sc GELATO} works better when the redshift of the host galaxy is known either from catalogs (e.g. the NASA/IPAC Extragalactic Database NED\footnote{http://ned.ipac.caltech.edu} or  HyperLeda\footnote{http://leda.univ-lyon1.fr}) or from direct measurement of narrow spectral features from the host. 

GELATO has recently received a number of improvements:
\begin{itemize}
\item the templates database was entirely revised. For the new templates, only spectra of very well studied SNe were hand-picked and visually examined one by one. SNe spectra with uncertain dating are not included. Thus, the best fitting templates now have clearly stated phases, which are relative to either explosion or maximum light dates obtained from the literature. At the moment of this writing, the templates database consists of more than 2200 spectra of 100 SNe of all types. New templates are being continuously added. 
\item new options for spectral noise reduction were added. Now the user can choose among various spectral smoothing or high frequency signal filtering algorithms. 
\item a reddeing correction can now be applied to the user spectrum before being compared with the templates.
\end{itemize}

Currently more enhancements to GELATO are being developed and will soon be available to the community. 

As expected, for spectra with fair-to-good S/N,  the three codes give very similar results. Infrequently, it happens that the results are contradictory or inconclusive, especially for SNe with very shallow spectral features. Therefore we always perform a visual validation of the code output that sometimes is also the only way to single out new peculiar events.

To each transient it is assigned the SN type of the best fit template. Following the consolidated classification scheme (e.g. Filippenko 1997, Turatto 2003), SNe are divided in the following classes:
\begin{description}
\item[{\bf Ia}:] their spectrum does not show H  but lines of intermediate mass elements in particular Si, Ca and S. SNe Ia are believed to result from the thermonuclear runaways of white dwarfs that reach the Chandrasekar limit after accreting material or merging with a close companion (Maoz at al. 2013). 
\item[{\bf II}:] they develop strong and broad H lines although at very early time they often exhibit an almost featureless blue continuum with shallow P-Cygni He features. They are the most frequent outcomes of the core collapse of massive stars (M$>8\,M_\odot$; Smartt 2009), including in this group the IIP and IIL.
\item[{\bf Ib/c}:] when the spectrum does not show H  or SiII lines, the SN is classified of type {\bf Ib} if it shows He features and otherwise of type {\bf Ic}. These events are thought to be the core-collapse of massive stars that during evolution have lost the H and for Ic also the He external layers (Clocchiatti \& Wheeler 1997). The evolutionary scenario leaves room for intermediate cases where only a thin He layer is left (Ib/c) or  also a thin H layer remains ({\bf IIb}).
\item[{\bf IIn}:]  they show intense narrow H$\alpha$ emissions, sometimes on top of broader components.  This feature is taken as a signature of interaction of the ejecta with a dense circumstellar medium (CSM). When the emission from the shocked region outshines the radiation from the SN \linebreak ejecta it can be difficult to assess the nature of the exploding star. The current understanding is that in majority type IIn results from the explosion of massive stars (Turatto et al. 1993) but in a few cases it has been argued that they occur after a type Ia thermonuclear explosion (Hamuy et al. 2003).
\end{description}

The standard SN types described above cover 95\% of the actual SN classifications.
However, a small number of peculiar events does not fit into this scheme. Among these events we mention the type~Ibn SNe (Pastorello et al. 2007, 2008a, 2008b), i.e. type~Ib/c SNe interacting with the CSM; super-luminous, i.e. extremely bright, SNe (SNSL, \linebreak Pastorello et al. 2010; Quimby et al. 2011);  SN impostors (Van Dyk et al. 2000, Maund et al. 2006, Van Dyk \& Matheson 2012), i.e. events which mimic the spectrum of a type IIn  SN but that are considered the super-outburst of a very massive luminous blue variable star, with the progenitor surviving the outburst. 
In all cases we report the designation of the best fit template SN(e) that allows for further implementation of different classification scheme.

After classification of a candidate, a formal communication is immediately sent to IAU and/or  ATEL for dissemination.
At the same time the fits spectrum and a snapshot of the best result of the cross-correlation with the archive of SN spectral templates are posted in  the Padova-Asiago SN group web pages. 

Figure~2 shows a few representative spectra of SNe classified during the ACP, a peculiar thermonuclear SN and four examples of core-collapse SNe. Going from the top to the bottom, it is shown the first official classification of the ACP, the peculiar type-Ia SN~2011fs (Tomasella et al. 2011a), similar to SN 1991T. 
SN~2011iw was classified soon after explosion by Tomasella et al. (2011b) as a test-case for the verification phase of the Gaia Science Alerts Working Group 
(GSAWG)\footnote{http://www.ast.cam.ac.uk/ioa/wikis/gsawgwiki}, aimed to verify the robustness of the alerts and to confirm them with a dedicated network of small follow-up telescopes.
SN~2012aw exploded in a close galaxy (Messier 95) and was promptly classified as a young type~II, about four days after the core collapse (Siviero et al. 2012). After classification, different competitive groups directly detected the red supergiant progenitor in Hubble Space Telescope images obtained several years before the explosion, as well as in ground-based, near-infrared images (Elias-Rosa et al. 2012; Van Dyk et al. 2012; Fraser et al. 2012; Van Dyk et al. 2013) and found out evidence for substantial circumstellar dust around the luminous star, whose absorptions were used as a case study by Kochanek et al. (2012). Several authors started a follow-up campaign of this close transient (Bayless et al. 2013; Bose et al. 2013; Dall'Ora et al. 2013), which went on till the nebular phase  (Jerkstrand et al. 2013).  \\
CSS121015:004244+132827 is to date the highest redshift transient observed during the ACP. Our first spectrum \linebreak (Tomasella et al. 2012a) shows a very blue, featureless continuum, in agreement with the previous description of Drake et al. (2012). A narrow absorption attributed to the host galaxy Mg II doublet (2796-2803\AA\/) and the detection of very weak Balmer emission lines (H${\beta}$, H${\gamma}$, H${\delta}$) allowed us to estimate the host galaxy redshift to be z=0.286. Adopting this value, the absolute magnitude of the transient was $-22.5$ mag, making it a bright member of the class of super-luminous, stripped envelope events, similar to SN 2010gx (Pastorello et al. 2010). An extensive monitoring campaign started soon after classification (Benetti et al. 2014).  \linebreak
SN~2013hc was discovered by a ten-years old boy\footnote{http://www.universetoday.com/107085/young-boys-discovery-confirmed-as-a-peculiar-supernova-explosion/} at the end of October. After a long period of bad weather conditions, we obtained a spectrum of this transient, under excellent seeing condition, when it had faded to about \linebreak $V\sim$20 mag. Even if the signal-to-noise ratio of this spectrum was around 15 (cf. Fig.~2),  it was possible 
to detect the presence of a low-contrast H${\alpha}$ emission features, from which we measured the ejecta velocity $v~$=3100 km s$^{-1}$, leading to the classification as a peculiar type~II supernova (Gray et al. 2013).

\section{Program performance}

\begin{figure*}
\begin{minipage}{.5\textwidth}
\includegraphics[width=\linewidth]{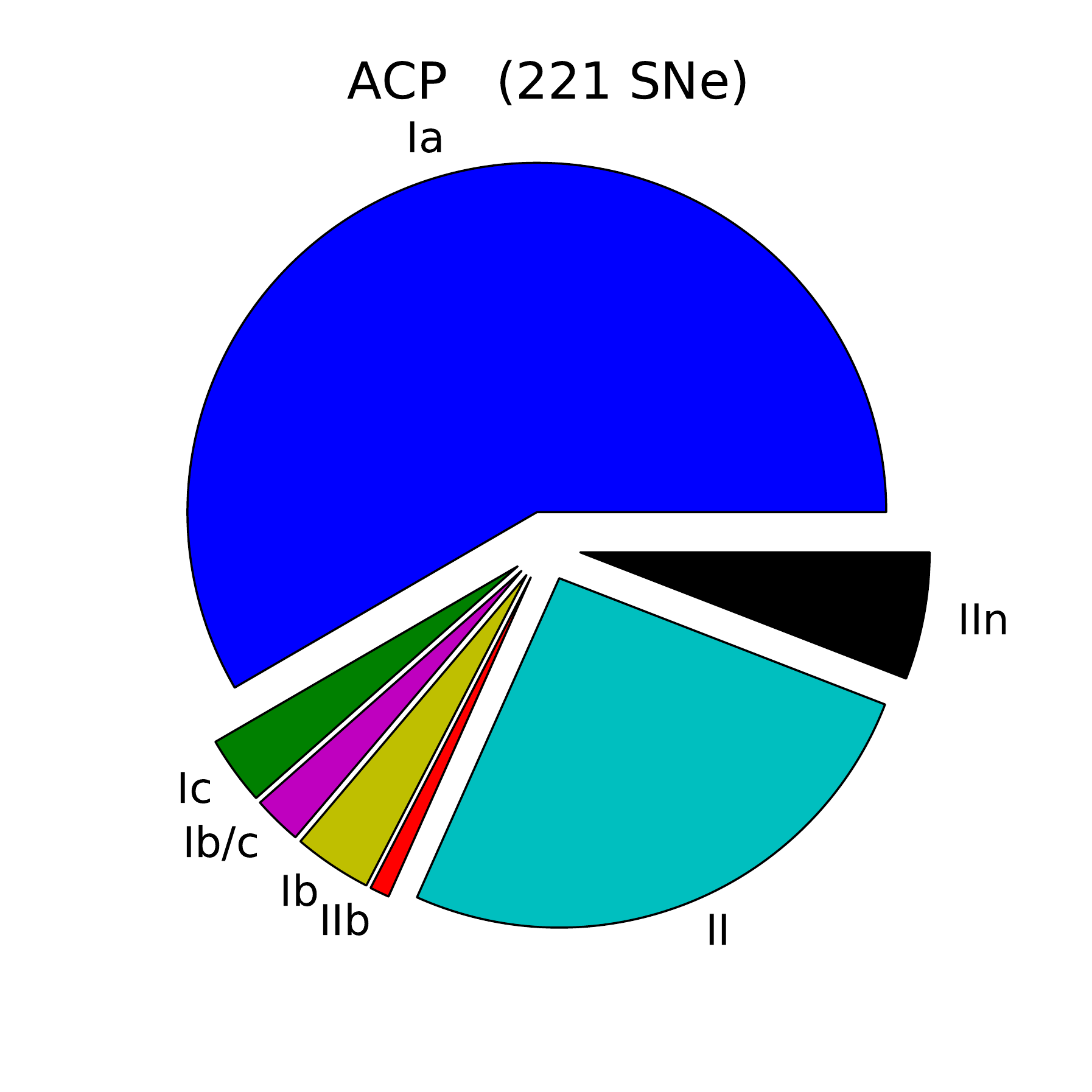}
\end{minipage}
\begin{minipage}{0.5\textwidth}
\includegraphics[width=\linewidth]{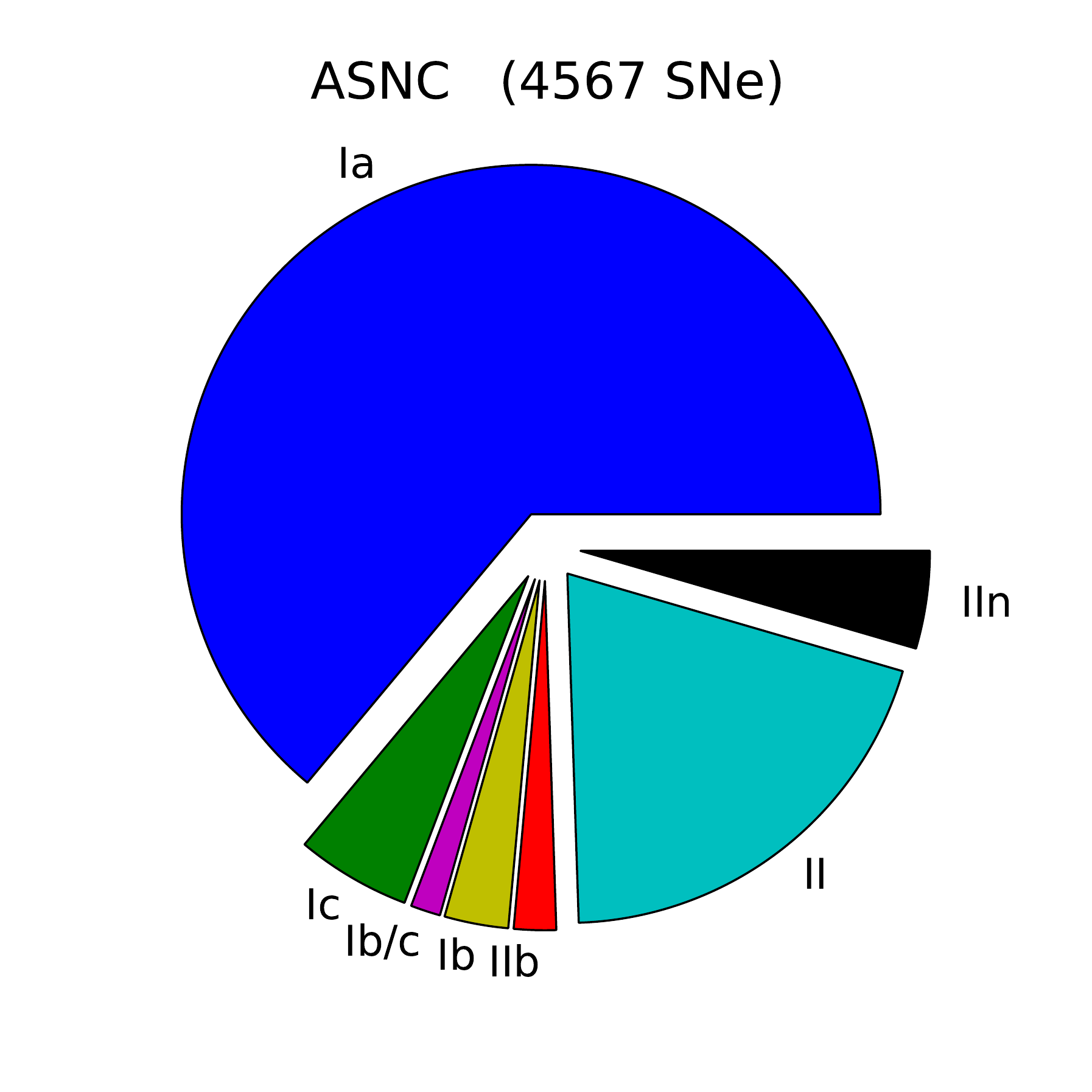}
\label{label1}
\end{minipage}
\caption{Left panel: Type distribution of the 221 SNe classified through the ACP (May 2011 - October 2013): 60.4\% type-Ia; 24.6\% type-II; 5.5\% type-IIn; 3.6\% type-Ib; 2.7\% type-Ib/c; 3.2\% type-Ic. Right panel: distribution of 4567 SNe collected in the ASNC (excluded the SNe with an uncertain classification), which lists all SNe with IAU designation announced via CBET.}
\end{figure*}

So far (May 2011 $-$ Oct. 2013), we spectroscopically confirmed and classified a total  of 221 SNe (the 	up-to-date full list is available in the Padova-Asiago Supernova Group web pages$^{11}$). 
An additional half a dozen of sources turned out to be of different nature, e.g. cataclysmic variable stars \linebreak (Tomasella et al. 2012b) or SN impostors (Tomasella et al. 2013b, Tomasella et al. 2013c, Dimai et al. 2013, Tartaglia et al. 2014). For another dozen of targets the spectral signal-to-noise ratio was too low to make definite assessments.

The distribution of SN types in our sample is shown in the left panel of Figure~3. Most SNe, 60\%,  are of type Ia; 25\%  are of type II and  5\% type IIn. The remaining events  are classified as type Ib (3.6\%) or Ic (3.2\%), with a fraction of intermediates or unclear cases labelled Ib/c (2.7\%). In each class we recognised a number of peculiar objects, e.g. faint type Ia  SNe similar to SN~1991bg (3), bright Ia similar to SN~1991T (4) and high velocity Ic (3).
Among the peculiars, we classified SN~2011hw (Valenti et al. 2011b) which is a transitional type IIn/Ib. 

The SN type distribution of our sample is compared with that of a general SN list of the ASNC$^8$. The catalog lists all SNe with IAU designation (now announced via CBETs). On December 2013 the catalog contained 6309 SNe. We recall that nowadays a major fraction of the SN discoveries are announced through ATel and do not receive a IAU standard designation. The ACP type distribution is very similar to that of the ASNC, hence we argue that our sample does not suffer from significant selection biases.

\begin{figure}
\includegraphics[width=\linewidth]{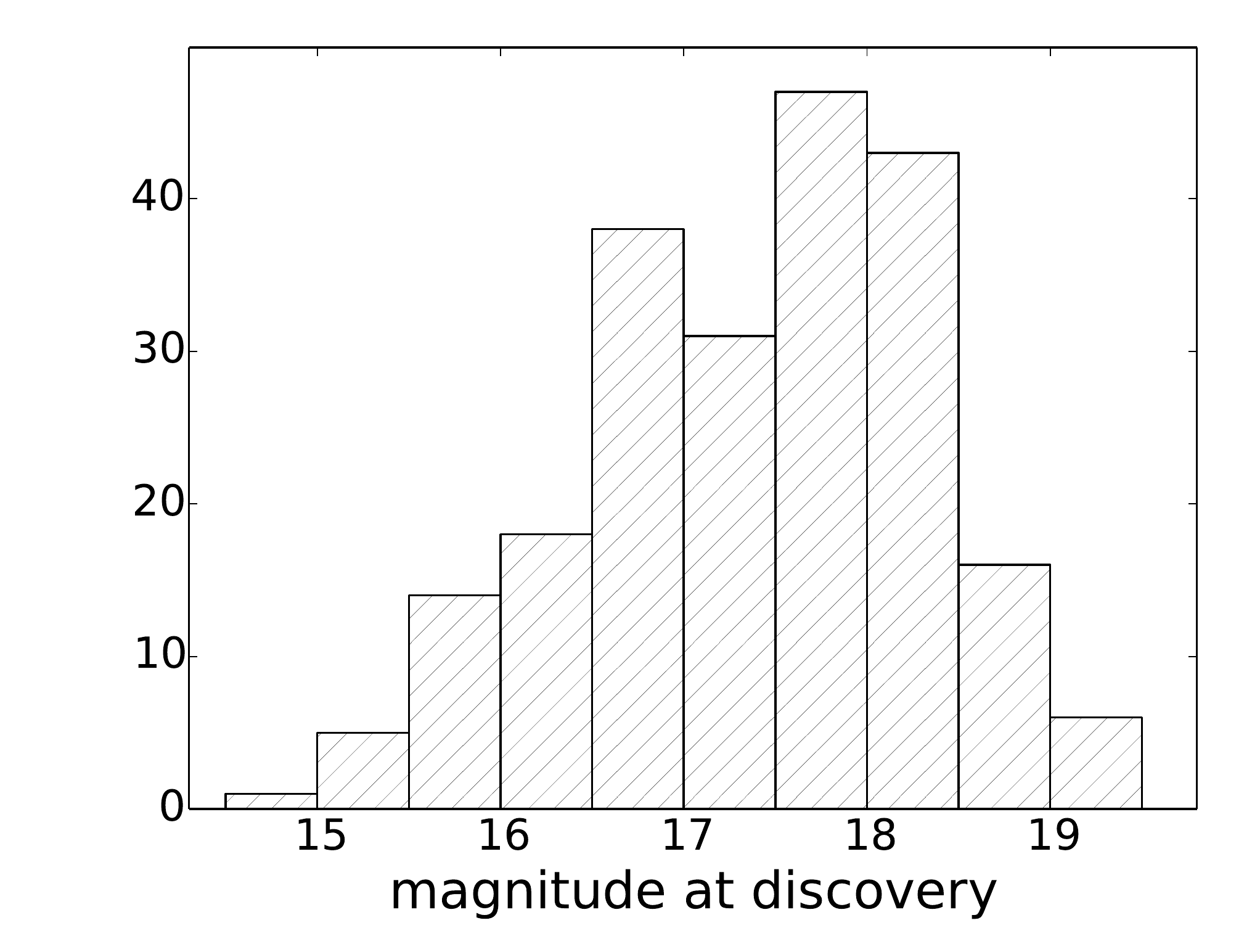}
\caption{Magnitude distribution of our sample of 221 SNe at discovery (unfiltered apparent magnitude). }
\label{magdist}
\end{figure}

\begin{figure}
\includegraphics[width=\linewidth]{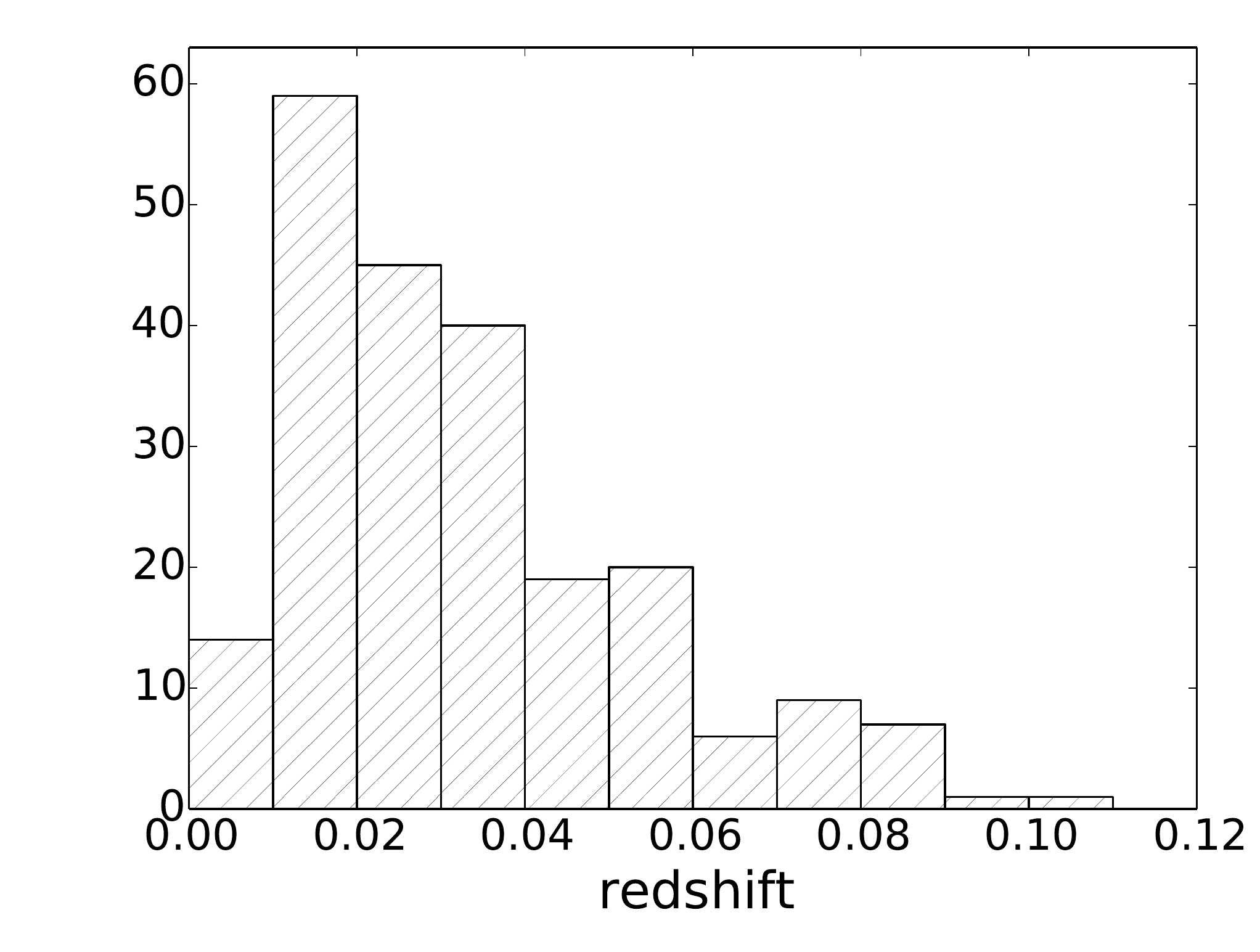}
\caption{Redshift distribution of our sample of 221 SNe.}
\label{redshift}
\end{figure}

Figure~\ref{magdist} shows the distribution of apparent magnitude of the new transients at discovery, as reported by the  discoverers  (typically unfiltered CCD magnitudes).  
The redshift distribution is plotted in Figure~\ref{redshift}. The mean redshift is $z=0.033$ with the most distant objects reaching $z=0.1$ (but the previously discussed superluminous type-II SN, shown in Figure~2, at z=0.286). For an Hubble constant \linebreak H$_0=73$ km s$^{-1}$ Mpc$^{-1}$, this corresponds to an average distance of 140 Mpc and to a distance limit of 400-500 Mpc. 

\begin{figure}
\includegraphics[width=\linewidth]{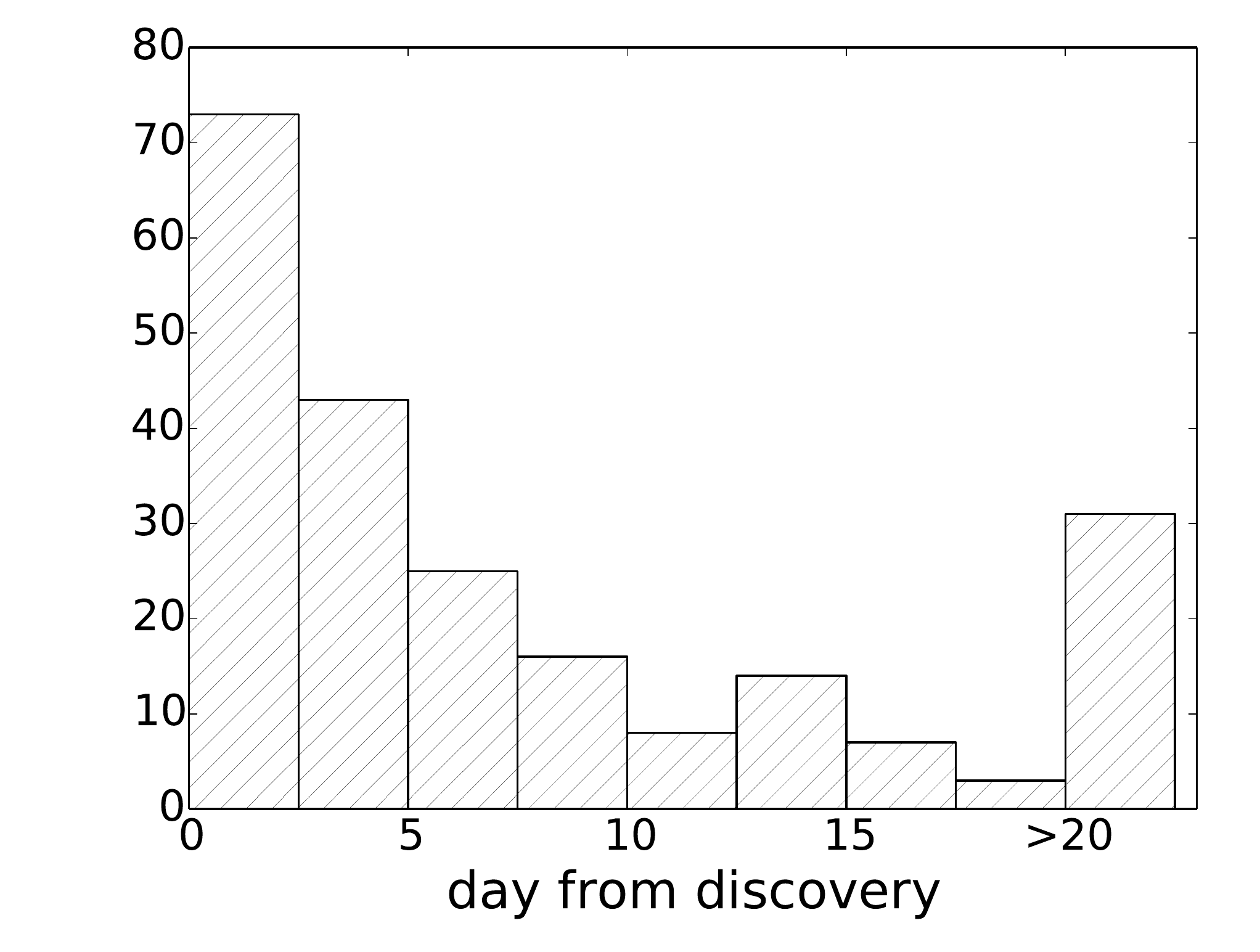}
\caption{Time interval in days between SN discovery and ACP spectral classification. }
\label{delay}
\end{figure}

An important test on the program performances is the delay from the SN detection to classification. This is shown in Figure~\ref{delay} where it appears that 50\% of the events  are classified within 5 days from detection and 25\% within 2 days. The delay is the sum of a number of components. First of all the discoverers need some time to process the search images, detect the candidates and post the discoveries. In some cases the discoverers chose to obtain a confirmation image before making a public report.
Our program is  based on scheduled nights and therefore there are observing gaps that can be even wider in case of bad weather.  To address these issues we update in real time our selected candidate list to include the last announced candidates. This guarantees that, within the limits of our observing schedule, we can maximize the number of very early classifications.

\begin{figure}
\includegraphics[width=\linewidth]{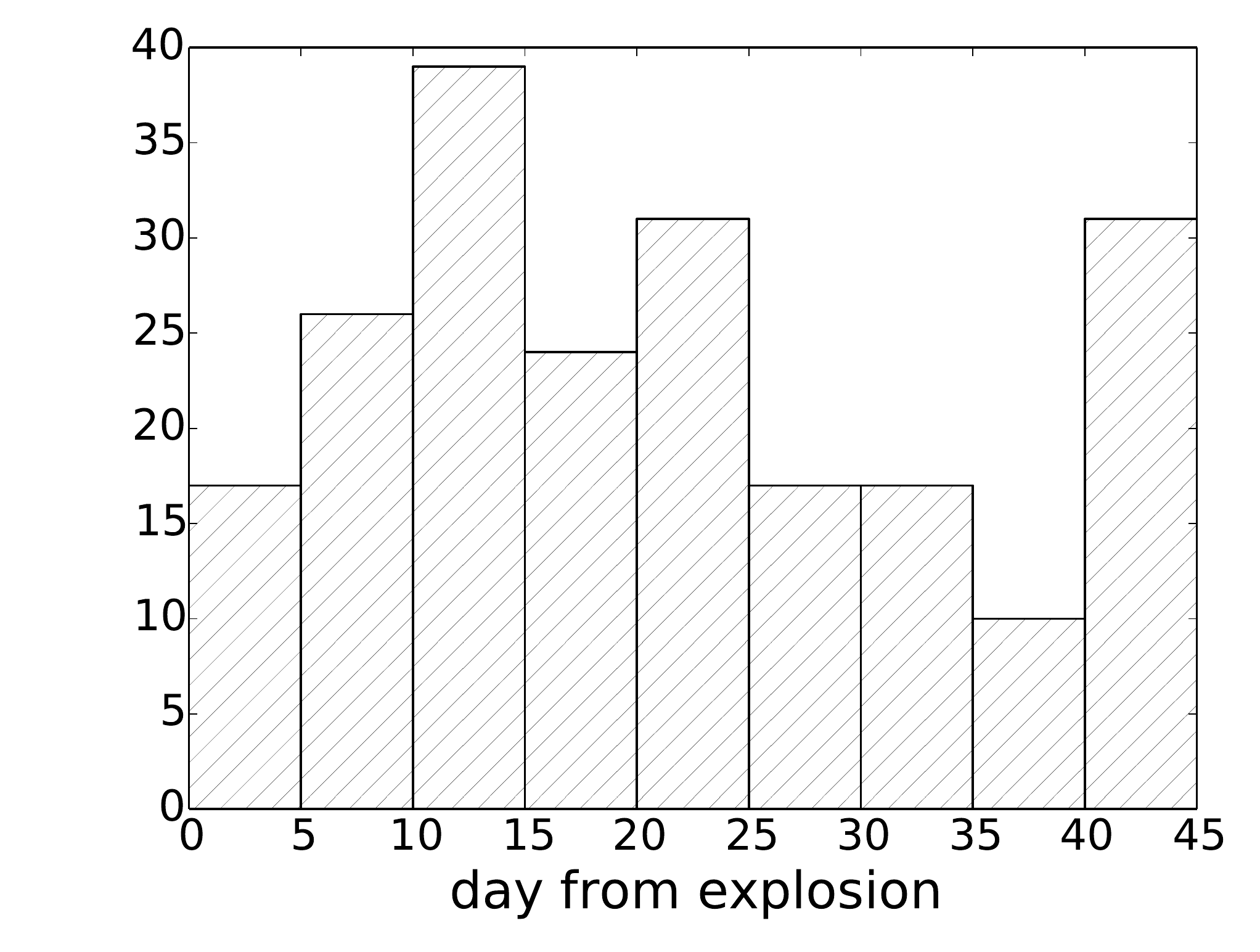}
\caption{Phase distribution of our sample of 221 SNe at the time of classification.}
\label{phase}
\end{figure}

The template-matching approach allows us for an estimate of the phase of the transient at the time of spectroscopic observation. The distribution of SN phases is shown in Figure~\ref{phase} where, for consistency among different SN types, we use as reference the epoch of explosion (for type-Ia we assume that this occurs 18 days before the day of the \linebreak  $B$-band maximum, cf. Folatelli et al. 2010). The distribution is fairly flat, with 50\% of the SNe discovered within 20 days from explosion.

 \section{Statistical analysis: host galaxy distribution} 
 
The statistical analysis of the distributions of classified SNe can lead to significant insights on the nature of the progenitors. In this respect, the observations and the classification criteria  should be as homogeneous as possible. From this perspective, the classified SNe of the ACP is expected to be valuable. In the following, we repeat the analysis on type~II SNe by Arcavi et al. (2010). They used the first compilation of 72 core-collapse SNe from the PTF to study the type distribution in dwarf and giant galaxies,  finding a significant excess of 
SNe~IIb in dwarf hosts and a concentration of type~Ib/c SNe in giant galaxies. 
We repeated the same analysis on our subset of core-collapse SNe. To this purpose, we retrieved the $r$-band absolute magnitudes of the host galaxies of our  SNe from the Sloan Digital Sky Survey (SDSS), Data Release 10\footnote{http://skyserver.sdss3.org/dr10/en/home.aspx} (Ahn et al. 2013), if available, or from NED. We removed the Galactic extinction, using Schlegel et al. (1998). 

\begin{figure}
\includegraphics[width=\linewidth]{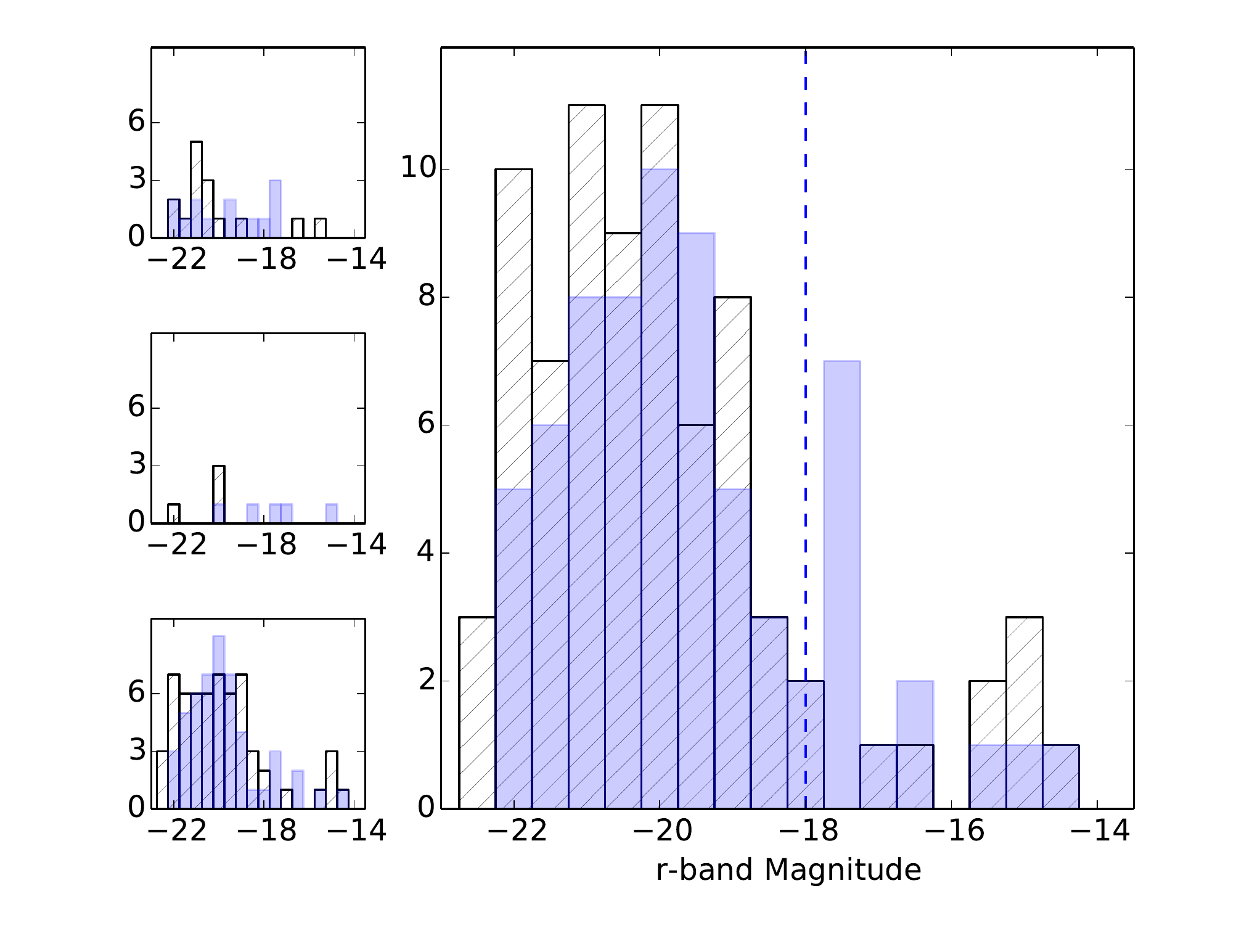}
\caption{Distribution of $r$-band absolute magnitude for the host galaxies of our sub-sample of 78 SNe (bold-line, hatched bars). We over-plot the 72 core-collapse SNe from Arcavi et al. 2010 (shaded bars, violet in the online journal). 
The dashed line at $M_{r}$=$-18$ mag  divides giant ($M_{r}<-18$ mag) from dwarf galaxies ($M_{r}\geq-18$ mag). 
Histograms for SN subtypes are presented on the left. From top to bottom:  SNe Ib/c; SNe IIb and SNe II. See text for details. }
\label{magabs}
\end{figure}

Our final sample consists of 78 core-collapse events, covering a wide range of host $r$-band absolute magnitudes. Our sample is therefore comparable to that of Arcavi et al. (2010). The histogram of host $r$-band absolute magnitudes is plotted in Figure~\ref{magabs}, hatched bars, compared to those from Arcavi et al. (2010), shaded bars. The dashed line at $M_{r}$=$-18$ mag  separates giant ($M_{r}<-18$ mag) from dwarf galaxies ($M_{r}\geq-18$ mag).  
We find that the overall distribution of the two samples are similar despite the fact that our SNe collection is heavily biased by targeted surveys (i.e. amateurs searches). The distribution of our collection of SNe peaks at $M_{r}$=$-21$ mag and a faint tail extends up to $M_{r}$=$-14$ mag. The larger number of core-collapse in giant with respect to dwarf galaxies is expected as effect of luminosity bias (cf. Cappellaro et al. 1988, 1993).   

Figure~\ref{magabs} shows also the distribution for different core-collapse sub-types: SNe Ib/c on the top, SNe IIb in the centre and SNe II on the bottom. 
Allowing for the possible ambiguities in the classification of stripped-envelope events with only a single epoch spectrum, we can say that, 
differently from Arcavi et al. (2010), SNe Ib/c are concentrated in giant hosts extending to dwarf galaxies, in the same way as type II and whole set of core-collapse SNe. Moreover, our 4 type IIb SNe are located only in luminous hosts, thus extending to the giant galaxies the distribution of Arcavi et al. (2010) type IIb sample. 

The speculation by Arcavi et al. (2010) of an absence of normal type Ic and of an excess of type IIb SNe in dwarf galaxies does not find confirmation in our sample. Our result is in  agreement with those of Sanders at al. (2012) who, based on spectroscopic metallicity measurements, found no statistical differences in the environments of Ib and Ic and between Ib and IIb.

 \section{Conclusions}
 
In this paper we have presented the Asiago Classification Program ACP and the results of the first two years of operation.

We show that a small telescope located in a site with continental weather conditions can give a significant contribution to the rapidly growing quest for transients classification. The positive experience of these two years encouraged us to further improve the telescope instrumentation and operational mode.
In particular, very recently, the re-furbishment of the telescope control system has been completed,  which is now used in remote mode. This allows the implementation of a new control room in the main buildings in Asiago, providing a quicker access to the 1.82m telescope and improving security for astronomers. We are now proceeding with the procurement of a new CCD 
(Andor IKON L936) with higher response in the UV and strongly  reduced fringing contamination red wards of 7500 \AA\/.

The quest for transients classification is going to grow even higher in the next years, in particular with the large number of transients that will be delivered by the ESA mission Gaia. 
Different simulations agrees that Gaia alone will detect over a thousand SNe per year (Cappellaro 2012, Altavilla et al. 2012, Blagorodnova et al. 2013). 
The survey strategy guarantees for an unbiased sky coverage and will allow to explore with better statistics many open science topics, such as for instance the relation between SN and host galaxy types. In fact, as we have shown in Sect.~5, existing data are still insufficient. 
The limiting magnitude of Gaia transients ($\sim19.5-20$) appears well suited for our instrumentation and therefore we plan to devote a special effort to the classification of these targets with the coordination of the Gaia Science Alerts Working Group.

\acknowledgements
Based on data obtained in Asiago, by the \linebreak Copernico 1.82m telescope operated by INAF OAPd, and the \linebreak Galileo 1.22m telescope operated by Dept. of Physics and Astronomy, University of Padova. 
 
L.T, S.B., E.C., A.P., M.T., and A.H. are partially supported by the PRIN-INAF 2011 with the project "Transient Universe: from ESO Large to PESSTO".
N.E.R. acknowledges the support from the European Union Seventh Framework Programme (FP7/2007-2013) under grant agreement n. 267251.

We acknowledge the INAF Centre for Astronomical Archives (IA2, http://ia2.oats.inaf.it/).   We used data from SDSS-III (Ahn et al. 2013).  
Funding for SDSS-III has been provided by the Alfred P. Sloan Foundation, the Participating Institutions, the National Science Foundation, and the U.S. Department of Energy Office of Science. This research has made use of the NASA/IPAC Extragalactic Database (NED) which is operated by the Jet Propulsion Laboratory, California Institute of Technology, under contract with the National Aeronautics and Space Administration. We acknowledge the usage of the HyperLeda database. We acknowledge the usage of the Astronomy Section of the Rochester Academy of Sciences (Latest Supernovae).

We thank the amateur astronomers for promptly communicating us their transients discoveries. 
We are grateful for excellent staff assistance at Asiago telescopes and for precious collaboration of our colleagues giving us the opportunity of taking some spectra in ToO during their observing runs.

%
%

\end{document}